\documentclass[aip, jcp, reprint, amsmath, amssymb, superscriptaddress]{revtex4-2}
\usepackage{graphicx,verbatim}
\usepackage{braket}
\usepackage{color}

\usepackage{soul}
\usepackage[normalem]{ulem}
\usepackage[x11names]{xcolor}
\usepackage{hyperref}
\hypersetup{colorlinks=true,linkcolor=blue}

\begin{document}
\bibliographystyle{apsrev4-2}

\title{Nonequilibrium Steady State Full Counting Statistics in the Noncrossing Approximation}

\author{Ido Zemach}
	\affiliation{
	School of Physics, Tel Aviv University, Tel Aviv 6997801, Israel
	}
 
\author{Andr\'e Erpenbeck}
\affiliation{
	Department of Physics, University of Michigan, Ann Arbor, Michigan 48109, USA
	}

\author{Emanuel Gull}
	\affiliation{
	Department of Physics, University of Michigan, Ann Arbor, Michigan 48109, USA
	}

\author{Guy Cohen}
	\affiliation{
	The Raymond and Beverley Sackler Center for Computational Molecular and Materials Science, Tel Aviv University, Tel Aviv 6997801, Israel
	}
	\affiliation{
	School of Chemistry, Tel Aviv University, Tel Aviv 6997801, Israel
	}

\date{\today}

\begin{abstract}

Quantum transport is often characterized not just by mean observables like the particle or energy current, but by their fluctuations and higher moments, which can act as detailed probes of the physical mechanisms at play. However, relatively few theoretical methods are able to access the full counting statistics (FCS) of transport processes through electronic junctions in strongly correlated regimes. While most experiments are concerned with the steady state properties, most accurate theoretical methods rely on computationally expensive propagation from a tractable initial state. Here, we propose a simple approach for computing the FCS through a junction directly at the steady state, utilizing the propagator noncrossing approximation (NCA). Compared to time propagation, our method offers reduced computational cost at the same level of approximation; but the idea can also be used within other approximations or as a basis for numerically exact techniques. We demonstrate the method's capabilities by investigating the impact of lead dimensionality on electronic transport in the nonequilibrium Anderson impurity model at the onset of Kondo physics. Our results reveal a distinct signature of one dimensional leads in the noise and Fano factor not present for other dimensionalities, showing the potential of FCS measurements as a probe of the environment surrounding a quantum dot.

\end{abstract}

\maketitle

\section{\label{sec:introduction}Introduction}

	Electronic transport through mesoscopic\cite{kouwenhoven_introduction_1997,datta_quantum_2005} and molecular\cite{nitzan_electron_2003,heath_molecular_2009,evers_advances_2020} junctions has long been a challenging subject for both experimental and theoretical research.
	Its study has contributed to our fundamental understanding of phenomena ranging from strong electronic correlations\cite{goldhaber-gordon_kondo_1998,cronenwett_tunable_1998,kouwenhoven_revival_2001,liang_kondo_2002} to interference effects in molecules.\cite{su_chemical_2016}
	It also has numerous potential device applications.\cite{datta_electronic_1990,coskun_high_2012}
    
    Most studies focus on the mean---or first moment---of the current.
    However, substantially more information can be gleaned from the higher moments of transport observables.\cite{landauer_noise_1998,esposito_nonequilibrium_2009}
    A well-known classical example of this is shot noise: by observing the fluctuations in a particle current, and considering the discrete nature of carriers, the charge of each carrier can be inferred.\cite{schottky_uber_1918}
    The noise---or second moment of the current---has also often been used in quantum analogs of this idea.\cite{de-picciotto_direct_1998,sela_fractional_2006,kumar_shot_2013,ferrier_universality_2016,kim_noise_2021,paoletta_determining_2024}
    Furthermore, the complete set of moments, which comprises the full counting statistics (FCS) of quantum transport observables, has often been of interest.\cite{levitov_charge_1993,levitov_electron_1996,bagrets_full_2002,levitov_counting_2004,gustavsson_counting_2006,tang_full-counting_2014,galperin_photonics_2017,honeychurch_full_2020,tang_full-counting_2017,landi_current_2024,kewming_first_2024,paulino_large_2024}
    For example, the generating function that contains all information regarding particle transport FCS can be used to obtain the first passage time of particle tunneling events,\cite{esposito_nonequilibrium_2009} and when combined with the FCS of energy transport it contains the full knowledge of the efficiency distribution characterizing thermoelectric quantum devices.\cite{verley_unlikely_2014, esposito_efficiency_2015,denzler_efficiency_2020,denzler_nonequilibrium_2024} 
    
   	The FCS of electronic transport is theoretically tractable in several important physical regimes of certain models, including the noninteracting,\cite{levitov_charge_1993,tang_full-counting_2014} Markovian\cite{bagrets_full_2002} and Toulouse point\cite{komnik_full_2005,schmidt_full_2007,gogolin_towards_2006} limits among others.\cite{larzul_energy_2023}
   	However, many physical regimes can only be reliably simulated with the aid of numerically exact methods.\cite{cohen_greens_2020}
   	In this category, some of the earliest studies employed the density matrix renormalization group,\cite{carr_full_2011,carr_full_2015} with later advances made by other matrix product state methods,\cite{guarnieri_full_2017,popovic_quantum_2021} iterative path integral techniques,\cite{nicolin_non-equilibrium_2011,simine_vibrational_2012,kilgour_path-integral_2019,gerry_full_2023} the hierarchical equations of motion method\cite{cerrillo_nonequilibrium_2016,schinabeck_hierarchical_2020} and inchworm Monte Carlo.\cite{ridley_numerically_2018,ridley_numerically_2019,ridley_lead_2019,pollock_reduced_2022}
   	
   	Other studies of the FCS rely on approximations.
   	When possible, this can grant access to predictions at a greatly reduced computational cost when compared to numerically exact methods, as well as in a larger class of models.
   	Most theoretical research in the field has been built around either master equations\cite{braggio_full_2006,flindt_counting_2008,esposito_self-consistent_2010,emary_counting_2009,schinabeck_current_2014,stegmann_detection_2015,kaasbjerg_full_2015,agarwalla_full_2015,kosov_waiting_2017,wang_unifying_2017,liu_frequency-dependent_2018,walldorf_noise_2020,stegmann_electron_2021,stegmann_higher-order_2022,mcconnell_strong_2022,brenes_particle_2023,kleinherbers_full_2023}  or nonequilibrium Green's functions techniques.\cite{avriller_electron-phonon_2009,schmidt_charge_2009,schmidt_charge_2009,haupt_current_2010,novotny_nonequilibrium_2011,park_self-consistent_2011,seoane_souto_dressed_2014,seoane_souto_transient_2015,agarwalla_full_2015,utsumi_full_2017,dong_full_2017,tang_short-time_2017,tang_heat_2018,tang_spin-resolved_2018,stadler_finite_2018,zhang_full_2020,davis_electronic_2021,seoane_souto_timescales_2020}
   	
   	In physical regimes where strong quantum correlations between the impurity and the bath orbitals are present, variations on the noncrossing approximation (NCA)\cite{keiter_perturbation_1970,bickers_review_1987} have been used in the literature to access shot noise.\cite{meir_shot_2002,wu_noise_2010}
   	However, approaches of this type have been difficult to apply to FCS.\cite{cohen_greens_2020}
   	A related approach is based on Hubbard Green's functions,\cite{miwa_towards_2017} and can provide reliable data at both the noninteracting and Markovian limits.
   	Recently, it was shown that propagator-based NCA methods can be formulated to the calculation of FCS in the presence of correlation effects, by simulating the dynamics starting from a separable initial state.\cite{erpenbeck_revealing_2021}
   	These resummed propagator hybridization expansions\cite{eckstein_nonequilibrium_2010} are the starting point for certain numerically exact quantum Monte Carlo methods,\cite{gull_continuous-time_2011,gull_numerically_2011,cohen_numerically_2013,cohen_greens_2014,cohen_greens_2014-1} and where needed can be efficiently evaluated to high orders using the inchworm Monte Carlo method.\cite{cohen_taming_2015,antipov_currents_2017,chen_inchworm_2017,chen_inchworm_2017-1,dong_quantum_2017,boag_inclusion-exclusion_2018,krivenko_dynamics_2019,eidelstein_multiorbital_2020,krivenko_dynamics_2019,eidelstein_multiorbital_2020,kleinhenz_dynamic_2020,kleinhenz_kondo_2022,kim_pseudoparticle_2022,strand_inchworm_2023,atanasova_stark_2024,goldberger_dynamical_2024}
   	
   	Here, we introduce a formulation of the propagator-based NCA method for the FCS of transport observables that works directly in the steady state, without the need to simulate relaxation dynamics.
   	Similar steady state methods have recently been used to explore correlation effects of normal observables both at the NCA level\cite{erpenbeck_resolving_2021} and the numerically exact level.\cite{erpenbeck_quantum_2023,erpenbeck_shaping_2023,kunzel_numerically_2024, erpenbeck_steadystate_2024}
   	However, as we discuss below, several challenges must be overcome to use these ideas for evaluating FCS.
   	The method captures some strong correlation effects and is computationally inexpensive, while having a straightforward generalization to more elaborate scenarios, such as models with both fermionic and bosonic baths.\cite{chen_anderson-holstein_2016}
   	Furthermore, the main ideas can also be extended to the inchworm Monte Carlo method, and thus have the potential to provide a highly efficient, numerically exact method for accessing FCS in the equilibrium or nonequilibrium steady state. 
    
    Using the method, we explore electronic transport within the single orbital Anderson Impurity Model (AIM) given different lead dimensionalities.
   This reveals the potential of local FCS transport measurements through an impurity as detailed probes of its environment, hinting at the possibility of future nanoelectronic device and sensor applications. 
    
    The outline of our paper is as follows: In Sec.~\ref{sec:model_and_FCS}, we introduce our model, discuss how different lead geometries can be described, and review the FCS framework.
    We introduce the NCA transport methodology in Sec.~\ref{sec:method}, where we also explain how the FCS can be calculated directly in the steady state.
    In Sec.~\ref{sec:Results}, we present our benchmarks for the steady state formulation of the FCS, along with an overview of the statistical observables for transport for different lead geometries.
    We summarize our work in Sec.~\ref{sec:conclusions}.

\section{Model and observables}\label{sec:model_and_FCS}
    \subsection{\label{sec:model}Anderson impurity model}
        We consider the single orbital AIM, a minimal description for a finite interacting quantum system coupled to an infinite noninteracting environment, such as a set of metallic leads.
        The Hamiltonian of the AIM consists of three summands:
        \begin{equation}
            H=H_{I}+H_{L}+H_{IL}. \label{eq:H_full}
        \end{equation}
        Here $H_{I}$ describes the interacting impurity, $H_{L}$ describes the environment, and $H_{IL}$ describes the coupling between the impurity and the environment.
    
        The impurity Hamiltonian is given by
        \begin{equation}
        H_{I}=\sum_{\sigma}\varepsilon_{\sigma}d_{\sigma}^{\dagger}d_{\sigma}+Ud_{\uparrow}^{\dagger}d_{\uparrow}d_{\downarrow}^{\dagger}d_{\downarrow}, \label{eq:H_I}
        \end{equation}
        where the $d_\sigma^{\left( \dagger\right)}$ denote creation/annihilation operators for an electron of spin $\sigma$ on the impurity orbital; $\epsilon_\sigma$ is the single particle occupation energy; and $U$ the Coulomb interaction strength.
        The single particle energies can be influenced by a capacitive gate voltage $V_{\text{gate}}$, which we model as $\varepsilon_{\sigma}=V_{\mathrm{\text{gate}}}-\frac{U}{2}$.
        
        The leads are described by a continuum of effective noninteracting states:
        \begin{equation}
         H_{L}=\sum_l \sum_{k\in l}\sum_{\sigma} \varepsilon_{\sigma k}a_{\sigma k}^{\dagger}a_{\sigma k}.
        \end{equation}
        Here, the $a_{\sigma k}^{(\dagger)}$ are creation/annihilation operators for a lead orbital indexed by $k$ and having spin $\sigma$, with associated energy $\varepsilon_{\sigma k}$.
        Orbital indices are grouped by their lead index $l$.
        To consider transport through a junction, we henceforth assume that the environment consists of two separate leads, referred to as the ``left'' and ``right'' leads and correspondingly denoted by $l \in \lbrace L,R \rbrace$.
        
        The coupling between the impurity and the leads is
        \begin{equation}
        H_{IL}= \sum_l \sum_{k\in l} \sum_{\sigma} (t_{\sigma k}a_{\sigma k}^{\dagger}d_{\sigma} + \text{hc.}),
        \end{equation}
        with coupling constants $t_{\sigma k}$.
        Rather than explicitly defining these parameters and the lead density of states, it is more convenient to express them in terms of a coupling density
        \begin{equation}
            \Gamma_{l}\left(\omega\right)=\pi \sum_\sigma \sum_{k\in l}\left|t_{\sigma k}\right|^{2}\delta\left(\omega-\varepsilon_{\sigma k}\right). \label{eq:coupling_density}
        \end{equation}
        When the leads are initially assumed to be in equilibrium, the observables we will be interested in can then be written in terms of the hybridization functions
        \begin{equation}
         \begin{aligned}
        	\Delta_{l}^{>}\left(t\right) &= \frac{1}{\pi}\int d\omega e^{i\omega t}\Gamma_{l}\left(\omega\right)f_{l}\left(\omega-\mu_l\right),\\
        	\Delta_{l}^{<}\left(t\right) &= \frac{1}{\pi}\int d\omega e^{-i\omega t}\Gamma_{l}\left(\omega\right)\left(1-f_{l}\left(\omega-\mu_l\right)\right). 
        \end{aligned}
        \label{eq:def_hyb}
        \end{equation}
        The latter depend on $f_l(\omega-\mu_l) \equiv  \frac{1}{1+\exp(\beta(\omega-\mu_l))}$, the Fermi occupation function characterizing lead $l$ in terms of the inverse temperature $\beta$ and chemical potential $\mu_l$.
        We model a bias voltage $V$ between the left and the right lead by symmetrically shifting the chemical potentials away from their equilibrium value, $\mu_L=-\mu_R=V/2$.
        Throught this work, we set $\hbar=k_\text{B}=e=1$.

    \subsection{\label{subsec:Coupling_Density}Describing different lead geometries}    

        An essential part of this work is to study the influence of different lead geometries and dimensionalities on transport properties.
        Here, we briefly outline how the hybridization function can be calculated for a given tight-binding model for the leads.
        For details we refer the reader to Ref.~\onlinecite{ridley_lead_2019}.

        Let $h$ be a single-particle Hamiltonian describing a specific lead geometry within the tight-binding approach.
        $h$ is written in the basis of a set of site orbitals labeled by the index $i$.
        The eigenvalues of $h$ are denoted by $\varepsilon_{\sigma k}$ and the associated eigenvectors by $\ket{k_\sigma}$, such that $h \ket{k_\sigma} = \varepsilon_{\sigma k} \ket{k_\sigma}$.
        The coupling between the impurity and the leads is usually assessed based on geometric considerations. We assume that the leads have a terminal site $i=\mathcal{T}$, which is the only site in the respective lead that couples to the impurity with strength $t_{\sigma \mathcal{T}}$.
        Under this assumption, Eq.~(\ref{eq:coupling_density}) can be written as
        \begin{equation}
            \Gamma_{l}\left(\omega\right)=\pi \sum_\sigma |t_{\sigma \mathcal{T}}|^2 \sum_{k\in l} |S_{\sigma k\mathcal{T}}|^2 \delta\left(\omega-\varepsilon_{\sigma k}\right) , \label{eq:coupling_density_trafo}
        \end{equation}
        where $S_{\sigma k\mathcal{T}} = \braket{k_\sigma|\mathcal{T}}$ are the overlap matrix elements between the energy eigenstates $\ket{k_\sigma}$ and the terminal site $\ket{\mathcal{T}}$. 
        The evaluation of Eq.~(\ref{eq:coupling_density_trafo}) can be carried out by diagonalizing $h$ numerically.
        This is, however, computationally impractical for accurately describing leads in the thermodynamic limit, which typically requires a large number of sites.
        We therefore use the kernel polynomial method,\cite{weisse_kernel_2006, ridley_lead_2019} where the coupling density $\Gamma_{l}\left(\omega\right)$ is expressed as a series of Chebyshev polynomials $T_n(\omega)$,
        \begin{eqnarray}
            \Gamma_{l}\left(\omega\right) &=& \pi |t_{\sigma \mathcal{T}}|^2 
                            \frac{1}{\pi\sqrt{1-\omega^{2}}}\left[\mu_{0}+2\sum_{n=1}^{\infty}\mu_{n}T_{n}\left(\omega\right)\right] .
        \end{eqnarray}
        The coefficients $\mu_n = \braket{\mathcal{T} | \alpha_n}$ are given by the action of the $n^\text{th}$ Chebyshev polynomial of the Hamiltonian $h$ on the terminal state, $\ket{\alpha_n} = T_n(h)\ket{\mathcal{T}}$.
        The latter is obtained from the recurrence relation $\ket{\alpha_{n+1}} = 2h \ket{\alpha_{n}} - \ket{\alpha_{n-1}}$ using a sequence of relatively inexpensive matrix--vector products.
        Convergence can be accelerated by convolving the result with an appropriate kernel, and we use the Lorentz kernel for this purpose.\cite{weisse_kernel_2006, ridley_lead_2019}

    \subsection{\label{subsec:FCS} Full counting statistics}  
        
        Here, we provide the basic definitions that will be used below to explore the FCS of particle transport.
        For more details, regarding the use of generating functions to describe FCS, we refer the reader to Ref.~\onlinecite{esposito_nonequilibrium_2009}.
        The evaluation of FCS within the real-time hybridization expansion is based on the mapping described in Ref.~\onlinecite{tang_full-counting_2014}, and a detailed account of all its technical aspects can be found in Refs.~\onlinecite{ridley_numerically_2018, ridley_numerically_2019, erpenbeck_revealing_2021, pollock_reduced_2022}.

        The central object we will seek to evaluate is the generating function of particle transport events, which is defined as
        \begin{eqnarray}
            Z_l\left(t,\lambda\right)	&=& \left\langle e^{i\lambda N_l}\right\rangle \left(t\right)
            = \text{\text{Tr}}\left\{ \rho_{\lambda}\left(t\right)\right\} .
            \label{eq:Genearating_function}
        \end{eqnarray}
        Here, $\lambda$ is the counting field, $N_l=\sum_\sigma\sum_{k\in l} a_k^\dagger a_k$ is the number of electrons in lead $l$, and $\rho(t)$ is the density operator of the system.  
        For the last equal sign, we have introduced the counting-field dressed Hamiltonian, $H_\lambda = e^{i\lambda N_l/2} H e^{-i\lambda N_l/2}$,
        as well as the dressed density operator $\rho_{\lambda}\left(t\right)=e{}^{-iH_{\lambda}t}\rho_{\lambda}\left(0\right)e^{iH_{-\lambda}t}$.

        While the generating function is a theoretical tool, its cumulants,
        \begin{equation}
            C_{k}(t)  = (-i)^k \frac{\partial^k}{\partial\lambda^k} \log \left(Z_l(t, \lambda) \right)\Big|_{\lambda=0} , \label{eq:def_cumulants}
        \end{equation}
        are tightly connected to physical observables.
        For the scope of this work, we are interested in the steady state current $I$ and noise $S$, which can be obtained by the time-derivatives of the first and second cumulant, 
        \begin{eqnarray}
            I &=& \lim_{t\rightarrow\infty} \frac{\partial}{\partial t} C_{1}(t), 
            \\
            S &=& \lim_{t\rightarrow\infty}\frac{\partial}{\partial t} C_{2}(t) .
        \end{eqnarray}
        Higher order observables and compound quantities such as the Fano factor $F=S/I$ can be obtained in a similar fashion once the generating function is known.

\section{Methodology\label{sec:method}}

    \subsection{\label{subse:NCA_HE} Hybridization expansion and NCA} 

        For the scope of this work, we use the NCA to calculate statistical observables associated with quantum transport.
        The name NCA has been used in the literature different contexts to describe several inequivalent methods that are perturbative in the impurity--lead coupling.\cite{keiter_perturbation_1970, grewe_diagrammatic_1981, kuramoto_self-consistent_1983, pruschke_anderson_1989, keiter_nca_1990, wingreen_anderson_1994-1, anders_beyond_1995,eckstein_nonequilibrium_2010}
        Here, we employ the NCA scheme based on the perturbative expansion of the restricted propagator (cf.\ Eq.~(\ref{eq:def_res_propagator})) in the impurity--environment coupling.
        This formulation of the NCA employs the stationary states of the decoupled impurity as a basis.
        Subsequently we outline how the FCS approach to transport is realized within the NCA, closely following Ref.~\onlinecite{erpenbeck_revealing_2021}.

        We will begin with the generating function $Z_l(\lambda, t)$.
        Assuming that the impurity and leads are decoupled at time $t=0$, the generating function can be expressed as
        \begin{equation}
        	Z_l(t, \lambda)	=	\sum_{\alpha\beta} \braket{\alpha|\rho_I|\alpha} K_{\alpha}^{\beta}(t,t,\lambda),  \label{eq:Z_NCA}
        \end{equation}
        with the vertex function
        \begin{eqnarray}
        	K_{\alpha}^{\beta}(t,t', \lambda)	&=&	\text{Tr}_L \left\lbrace \rho_L
        								\bra{\alpha} U_{-\lambda}^\dagger(t) \ket{\beta}\bra{\beta} U_\lambda(t') \ket{\alpha}
        								\right\rbrace . \nonumber \\ \label{eq:def:K_chi} 
        \end{eqnarray}	
        Here, $\alpha$ and $\beta$ are impurity states; $\rho_L$ and $\rho_I$, respectively, are the density operators of the leads and impurity at time $t=0$; and $\text{Tr}_L$ denotes the trace over all environment degrees of freedom.
        We have also introduced the counting-field-dressed time evolution operator, 
        \begin{equation}
            U_{\pm\lambda}(t) \equiv \mathrm{T}\exp\left(-i\int_0^t H_{\pm\lambda}(\tau) d\tau\right).
        \end{equation}

        The NCA is a hybridization expansion, meaning that it relies on a perturbative expansion of Eq.~(\ref{eq:def:K_chi}) in the coupling $H_{IL}$ between the impurity and the environment.
        Specifically, the NCA represents a resummation scheme accounting for an infinite subclass of all possible terms within the hybridization expansion, which is obtained by treating the lowest nonvanishing contribution to the hybridization expansion for the vertex function self-consistently. 
        The corresponding expression for the vertex function can be cast into the form of a Dyson equation, 
        \begin{eqnarray}
            \label{eq:Dyson_K}
            K_{\alpha}^{\beta}(t,t',\lambda)
            &=&
            k_{\alpha}^{\beta}\left(t,t^{\prime}\right)
            +\sum_{\alpha^{\prime}\beta^{\prime}}\int\limits _{0}^{t}\int\limits _{0}^{t^{\prime}}d\tau_{1}d\tau_{1}^{\prime} \times \\ && \nonumber \hspace{-1cm}
            k_{\beta'}^{\beta}\left(t-\tau_{1},t'-\tau_{1}'\right)
            \xi_{\alpha'}^{\beta'}\left(\tau_{1}-\tau_{1}', \lambda\right)K_{\alpha}^{\alpha'}\left(\tau_{1},\tau_{1}',\lambda\right),
        \end{eqnarray}
        with the contribution of the cross-branch hybridization given by
        \begin{eqnarray}
        	\xi_{\alpha}^{\beta}(t, \lambda)
        		&=&
        			\sum_{\sigma}
        			\sum_{l}
        					\Big( 
        					\Delta_{l}^<(t)
        					e^{-i\lambda t}
        					\braket{\alpha|d_\sigma|\beta} \braket{\beta|d_\sigma^\dagger|\alpha}
        					\nonumber \\ &&
        					+\Delta_{l}^>(t)
        					e^{i\lambda t}
        					\braket{\alpha|d_\sigma^\dagger|\beta} \braket{\beta|d_\sigma|\alpha}
        					\Big).
        \end{eqnarray}
        The name ``noncrossing approximation" refers to the fact that the terms of this equation can be represented by diagrams where the hybridization lines do not cross each other in time (see e.g.\ Ref.~\onlinecite{cohen_greens_2014}).
        Extensions to this approach can then be constructed by including higher-order corrections.\cite{pruschke_anderson_1989, haule_anderson_2001, cohen_greens_2014}

        The zeroth-order approximation for the vertex function, which corresponds to the first term in Eq.~(\ref{eq:Dyson_K}), can be written as
        \begin{equation}
            k_{\alpha}^{\beta}(t,t') = \delta_{\alpha\beta} G_{\alpha}^*(t)G_{\beta}(t'),
        \end{equation}
        with the single-branch restricted propagator
        \begin{equation}
        	G_{\alpha}(t) = \braket{\alpha |\text{Tr}_L \left( \rho_L U_\lambda(t) \right) | \alpha}  .\label{eq:def_res_propagator}
        \end{equation}
        The single-branch restricted propagator is also treated perturbatively in the impurity--environment coupling.
        Similarly to the vertex function, $G_{\alpha}(t)$ is determined by imposing a Dyson-like self-consistency condition based on the lowest order expression for the single-branch restricted propagator,
        \begin{eqnarray}
            \label{eq:Dyson_G}
            G_{\alpha}(t) &=& g_{\alpha}(t)\\ && \nonumber
             -\int_{0}^{t}\int_{0}^{\tau_{1}}d\tau_{1}d\tau_{2}g_{\alpha}\left(t-\tau_{1}\right)\Sigma_{\alpha}\left(\tau_{1}-\tau_{2}\right)G_{\alpha}\left(\tau_{2}\right),
        \end{eqnarray}
        with the single-contour self-energy 
        \begin{eqnarray}
        	\Sigma_{\alpha}(t)	&=&	
        	\sum_{\sigma}
        	\sum_{l}
        	\sum_{\beta} \Big( 
        				\Delta_{l}^<(t) \cdot \braket{\alpha | d_\sigma | \beta} \braket{\beta | d_\sigma^\dagger | \alpha} \nonumber \\&&
        				+  \Delta_{l}^>(t) \cdot \braket{\alpha | d_\sigma^\dagger | \beta}  \braket{\beta | d_\sigma | \alpha} \Big) \cdot G_{\beta}(t) . \label{eq:def_Sigma_G}
        \end{eqnarray}
        Finally, $ g_{\alpha}(t)=e^{-iH_It}$.
        We note that $G_\alpha(t)$ remains unmodified by the counting field, as all counting-field dependency cancels out on a single Keldysh branch.
        For the diagrammatic representation of this expansion, we refer to Refs.~\onlinecite{cohen_greens_2014, chen_anderson--holstein_2016}.

        To conclude our overview of the NCA approach, we comment on the applicability of this methodology.
        As the NCA is a low order self-consistent method that is perturbative in the impurity--environment coupling, it is best suited for the description of the strong interaction regime.
        Due to its nonlinear nature, however, it is hard to assess its regimes of validity rigorously, a problem that is exacerbated under nonequilibrium conditions.
        Previous works indicated that the NCA as presented here can provide qualitatively accurate data for temperatures that are not too far below the Kondo temperature,\cite{eckstein_nonequilibrium_2010} as well as in the large $U$ limit and for small bias voltages.\cite{cohen_numerically_2013}
        Yet, the NCA does not capture the correct behavior in the Kondo scaling regime, where vertex corrections become important\cite{anders_beyond_1995, anders_perturbational_1994, grewe_conserving_2008, eckstein_nonequilibrium_2010} and methods based on the renormalization group approach are more suitable.\cite{gerace_low-temperature_2002, kroha_conserving_2005}
        In this work, we therefore restrict our investigations to the high-energy remnants of Kondo physics and nonequilibrium effects.
        Still, the results presented in Sec.\ \ref{sec:results} should be considered qualitative rather than quantitative until verified by a more accurate methodology.

    \subsection{Full counting statistics in the steady state \label{sec:SS-NCA}}
        The defining property of the steady state is that all physical observables are constant in time.
        The generating function of transport event FCS is not a physical observable, but the time derivatives of its moments are.
        The fact that at steady state the cumulants characterizing electronic transport are constant in time implies that
        \begin{equation}
                \lim_{t\rightarrow\infty} \left( \frac{\partial^2}{\partial t^2} \left(\frac{\partial^{k}}{\partial\lambda^{k}}\log\left(Z_l(t, \lambda)\right) \Big|_{\lambda=0} \right) \right)	= 0.
                \label{eq:FCS scaling generating function}
        \end{equation}
        This realization suggests an asymptotic form for the generating function $Z_{l\ ss}\left(t, \lambda \right)$ at the steady state limit:
        \begin{equation}
            Z_l(t, \lambda) \xrightarrow{t \rightarrow \infty} Z_{l}^\text{ss}(\lambda) \cdot e^{w(\lambda)t}.\label{eq:asymptotic_ansatz}
        \end{equation}
        Here, the entire dependence of time derivatives on the counting field, and hence information about observables, is absorbed into the (complex) scaling function $w(\lambda)$. By inserting the steady state generating function into Eq.~\eqref{eq:def_cumulants}, we arrive at an expression for the time derivatives of all cumulants in the steady state:  
        \begin{equation}
            \frac{\partial}{\partial t}C_{k}(t)  = (-i)^k \frac{\partial^k}{\partial\lambda^k}w(\lambda)\Big|_{\lambda=0}.\label{eq:def_cumulants_steady_state}
        \end{equation}
        
        Eq.~\eqref{eq:asymptotic_ansatz} has implications for the form of the vertex function.
        Leveraging the relation between the generating function and the vertex function, Eq.~(\ref{eq:Z_NCA}), together with the assumption that the steady state vertex function cannot depend on the initial state or explicit times,\cite{erpenbeck_resolving_2021,erpenbeck_quantum_2023} we arrive at an ansatz for the counting-field-dependent vertex function in the steady state,
        \begin{equation}
            K^\beta_\alpha(t,t',\lambda) \xrightarrow{t,t' \rightarrow \infty} 
            K^\beta_{\text{ss}}(\Delta t, \lambda) \cdot e^{w(\lambda)T}.\label{eq:ansatz_ss_K}
        \end{equation}
        This is written in terms of the relative time $\Delta t=t-t'$ and the average time $T=(t+t')/2$.
        Inserting the ansatz Eq.~(\ref{eq:ansatz_ss_K}) for the steady state vertex function into Eq.~(\ref{eq:Dyson_K}),
        the exponential factor $e^{w(\lambda)T}$ from Eq.~(\ref{eq:ansatz_ss_K}) appears on both sides of the equation  and hence cancels out.
        This allows us to take the infinite time limit $T \rightarrow \infty$, where the result is expected to be independent of the initial condition.
        Given this, we arrive at
        \begin{equation}
			\begin{aligned}K_\text{ss}^{\beta}\left(\Delta t,\lambda\right) & =\int_{-T}^{\Delta t}\mathrm{d}\tau_{1}\int_{\tau_{1}}^{T}\mathrm{d}\Delta\tau \thinspace G_{\beta}^{*}\left(t-\tau_{1}\right)\\
				& \times G_{\beta}\left(\Delta\tau-\tau_{1}\right)\sum_{\gamma}\xi_{\gamma}^{\beta}\left(\Delta\tau,\lambda\right)\\
				& \times K_\text{ss}^{\gamma}\left(\Delta\tau,\lambda\right)e^{w\left(\lambda\right)\left(\tau_{1}-\Delta\tau\right)}.
			\end{aligned}
		\label{eq:SS_vertex}
        \end{equation}
        This is the defining equation for $K_\text{ss}^{\beta}(t, \lambda)$.
        As Eq.~(\ref{eq:SS_vertex}) does not have an inhomogeneous part, solutions are unbound with respect to multiplication by a constant. 
        Without loss of generality, we impose the normalization constraint $\sum_{\beta} K_\text{ss}^{\beta}(0,\lambda)= e^{w(\lambda)T}$, which is consistent with the normalization of the population probability when $\lambda=0$.
        The parameter $T$ then returns to the calculation, but can be chosen as arbitrarily large so that the condition $K^\beta_\alpha(T,T,\lambda) = K^\beta_{\text{ss}}(0, \lambda)$ is obeyed to within the desired numerical accuracy.

        As a last step, we derive an equation for the scaling function $w(\lambda)$. To this end, we use the ansatz from Eq.~(\ref{eq:ansatz_ss_K}) to relate the steady state vertex function to its counterpart at slightly different times:
        \begin{eqnarray}
            K^\beta_\alpha(t,t, \lambda) &=&
            K^\beta_\alpha(t-\Delta s,t-\Delta s, \lambda) e^{w(\lambda)\Delta s} . \label{eq:connection_vertex}
        \end{eqnarray}
        While this holds for any time $t$ in the steady state regime, it is insightful to consider the time $T$ used for the normalization of the steady state vertex function.
        In this case, we can identify $K^\beta_\alpha(T,T,\lambda) = K^\beta_{\text{ss}}(0, \lambda)$ on the left hand side of Eq.~(\ref{eq:connection_vertex}), and determine the scaling function according to 
        \begin{eqnarray}
             w(\lambda)  &=&
            \log\left(\frac{K^\beta_{\text{ss}}(0, \lambda)}{K^\beta(T^-, T^-, \lambda)}\right) / \Delta s ,
            \label{eq:w}
        \end{eqnarray}
        with $T^- = T - \Delta s$.
        We calculate the vertex function in the denominator from
        \begin{equation}
			\begin{aligned}K^{\beta}\left(T^{-},T^{-},\lambda\right) & =\sum_{\gamma}\int\limits_{0}^{T^{-}}\mathrm{d}\tau\int\limits_{\tau-T^{-}}^{\tau}\mathrm{d}\Delta\tau\thinspace G_{\beta}^{*}\left(T^{-}-\tau\right)\\
				& \times G_{\beta}\left(T^{-}+\Delta\tau-\tau\right)\xi_{\gamma}^{\beta}\left(\Delta\tau,\lambda\right)\\
				& \times K_\text{ss}^{\gamma}\left(\Delta\tau,\lambda\right)e^{w(\lambda)\left(\tau-\frac{\Delta\tau}{2}\right)}.
				\label{eq:prop_K_special}
			\end{aligned}
        \end{equation}
        Eq.~(\ref{eq:prop_K_special}) corresponds to the time propagation of the vertex function in Eq.~(\ref{eq:Dyson_K}), where the initial condition has been neglected and the vertex function on the right hand side has been replaced by its steady state counterpart.
        As this effectively neglects the details of the transient dynamics, the computational scaling of evaluating Eq.~(\ref{eq:prop_K_special}) is more favorable than performing a full time propagation; the effective scaling of the steady state scheme is linear in $T$.
        
        Given an initial guess for $w(\lambda)$ and $K_\text{ss}^{\beta}\left(\Delta t,\lambda\right)$, Eqs.~(\ref{eq:SS_vertex}), (\ref{eq:w}), and (\ref{eq:prop_K_special}) embody a self-consistent procedure for obtaining new guesses.
        This is the main element of our method.

    \subsection{Numerical scheme and scaling}\label{sec:scaling}

        At this point, it is useful to discuss the numerical cost associated with different ways to calculate transport FCS within the NCA method.
        If the counting-field-dressed vertex function is calculated by time propagation using Eq.~(\ref{eq:Dyson_K}), the computational cost scales like a standard NCA calculation.\cite{erpenbeck_revealing_2021}
        Assuming a time-independent Hamiltonian so that convolutions can be evaluated with the aid of a fast Fourier transforms, this implies a scaling of $O\left(N^2\log(N)\right)$, where $N$ is the number of points on the time grid.\cite{eckstein_nonequilibrium_2010}
        This often makes it difficult to access steady state physics, especially in the presence of slow dynamics.\cite{nordlander_how_1999}

        The considerations outlined in Sec.\ \ref{sec:SS-NCA} facilitate the calculation of the FCS directly in the steady state. 
        We start by calculating the single-branch restricted propagator $G_\alpha(t)$ using Eq.~(\ref{eq:Dyson_G}), which is independent of the counting field. 
        Assuming an initial guess for the scaling function $w(\lambda)$, we proceed as follows:
        \begin{enumerate}
            \item Calculate $K^\beta_{\text{ss}}(\Delta t, \lambda)$ by iterating Eq.~(\ref{eq:SS_vertex}) until self-consistency.
            \item Calculate $K^\beta(T^-, T^-, \lambda)$ using Eq.~(\ref{eq:prop_K_special}) and the current estimate for $K^\beta_{\text{ss}}(\Delta t, \lambda)$, which was obtained in step 1. 
            \item Calculate a new estimate for $w(\lambda)$ using Eq.~(\ref{eq:w}).
        \end{enumerate}
        These three steps are then repeated until convergence is achieved.
        Typically, we have found that the number of iterations needed to achieve self-consistency is of the order of ten.
        The bottlenecks in this scheme, which set the overall scaling, are steps 1 and 2; both of these scale as $O\left(N\log{N}\right)$, a linear advantage when compared to time propagation.
        In Appendix \ref{app:scaling} we present a numerical verification of this scaling behavior for a representative example.

\section{\label{sec:Results}Results}
    \begin{figure}[tb]
        \includegraphics[width=0.5\textwidth]{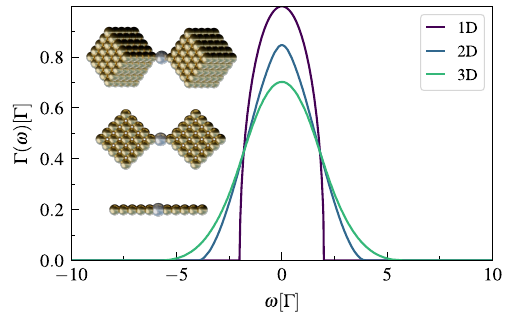}
        \caption{Coupling density and illustration of the geometry for different lead dimensionalities.
        The hopping parameters are set to $t_{\sigma \mathcal{T}}= t_{\text{tb}} =\Gamma$.
        \label{fig:coupling_density_and_systems}}
    \end{figure}

    We now continue to an application of our method to steady state FCS in the nonequilibrium AIM.
    In particular, we study the influence of lead dimensionality on the particle current and noise, by considering three distinct cases.
    In the first, an impurity is attached to two 1D chains; in the second, to the corners of two 2D square lattice sheets; and in the third, to the corners of two 3D cubic lattice cubes.
    These three geometries are described by respective noninteracting tight binding models with hopping parameters $t_{\text{tb}}$.
    They are illustrated in Fig.~\ref{fig:coupling_density_and_systems}, which also shows the coupling density associated with each case.
    The latter are obtained by the approach outlined in Sec.~\ref{subsec:Coupling_Density}.
    We note that a similar figure has been presented in Ref.~\onlinecite{ridley_lead_2019}, but that work---which relied on substantially more computationally expensive methods---continued to analyze only the 1D case, and focused on high temperature regimes.
    
    The 1D coupling density can be evaluated analytically.\cite{newns_self-consistent_1969, cuevas_molecular_2010, ridley_lead_2019}
    The result is given by
    \begin{equation}
    	\Gamma_l(\omega) = \frac{t_{\sigma \mathcal{T}}^2}{2 t_{\text{tb}}^2} \sqrt{4t_{\text{tb}}^2-(\omega-\mu_l)^2} \cdot \Theta(2t_{\text{tb}}-|\omega-\mu_l|),
    \end{equation}
    where $\Theta$ is the Heaviside step function.
    This coupling density has a maximal value of $\Gamma \equiv \frac{t_{\sigma \mathcal{T}}^2}{t_{\text{tb}}}$, which we employ as our energy unit throughout the rest of this work.
    
    In the 1D chain the total bandwidth of $W=4t_{\text{tb}}$ is set by two sharply defined edges.
    When the lead's dimensionality is increased,  the peak is lowered and the edge becomes smeared out (see Fig.~\ref{fig:coupling_density_and_systems}). In the rest of the paper, we will consider the 1D bandwidth as a rough measure of the bandwidth in other dimensions.
    
    Apart from the description of the leads, we choose the parameters of our model to be
    $U=8\Gamma$ and $\epsilon_\sigma = V_{\text{gate}} - \frac{U}{2}$, where the particle--hole symmetric case is realized for $V_{\text{gate}}=0$.
    The interaction $U$ drives the system to strongly correlated states at sufficiently low temperatures.
    While the precise temperature for the onset of correlation effects depends on the details of the leads, a rough estimate based on Bethe ansatz \cite{hewson_kondo_1997} in the case of 1D equilibrium case suggest the onset of correlations below a temperature of $T_{K} \sim 0.87 \Gamma$.
    We note that the NCA is not expected to reproduce the exact Kondo temperature and provides at most a qualitative picture. 
    Nevertheless, we found that this value is reasonably consistent with the temperature at which a resonance in the conductance appears.
    In the following, we will follow previous works\cite{goldhaber-gordon_kondo_1998, erpenbeck_resolving_2021} and use the increase in conductance as our operational definition for the onset of correlated physics.
    
    Because of the limitations of the NCA, we restrict our analysis to the edge of the correlated regime rather than going deeply into the Kondo scaling regime.
    The NCA does not properly reproduce the correct scaling behavior, the treatment of which requires more accurate methods.

    \subsection{\label{sec:benchmark} Benchmarking the steady state formulation}

		\paragraph*{General setup.}
        We begin by benchmarking the methodology presented in Sec.~\ref{sec:method}, which obtains the FCS directly in the steady state, against data obtained from the previously established technique relying on simulation of the full dynamics by way of time propagation.
        As a test case we choose to examine the 1D system (see Fig.~\ref{fig:coupling_density_and_systems}) at temperature $T=0.5\Gamma$ and $t_{\text{tb}}= 4\Gamma$ and $t_{\sigma \mathcal{T}}= 2\Gamma$, such that the bandwidth is $W=16\Gamma$.
        We consider two gate voltages: $V_{\mathrm{gate}}=0$, where the system is particle--hole symmetric; and $V_{\mathrm{gate}}=4\Gamma$,  where this symmetry is broken and the singly occupied states are at the same energy as the unoccupied state.
        
        \begin{figure}
            \includegraphics{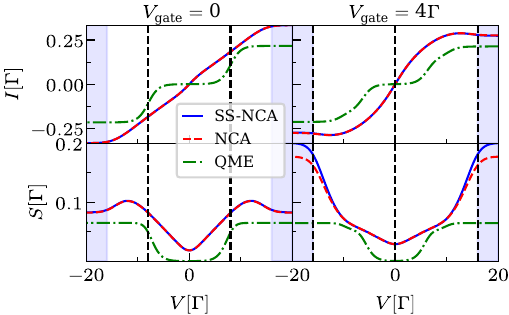}
            \caption{
                Current (top panels) and noise (bottom panels) extracted from the FCS for a 1D lead.
                Two  gate voltages are shown: $V_{\mathrm{gate}}=0$ (left panels) and $V_{\mathrm{gate}}=4\Gamma$ (right panels).
                The vertical dashed lines mark the condition for the onset of resonant transport, $\pm V/2=\epsilon_{\sigma}$ and $\pm V/2 = \epsilon_{\sigma}+U$.
                The blue shaded areas mark bias voltages that exceed the bandwidth.
                See text for other parameters.
                \label{fig:benchmark} 
            }
        \end{figure}

		\paragraph*{Comparison with time propagation.}
        Fig.~\ref{fig:benchmark} shows the steady state current and noise as a function of the bias voltage.
        Results from the steady state FCS framework (SS-NCA) are plotted alongside the corresponding time-dependent formulation (NCA) at time $t=10\Gamma$.
        To obtain some insight, we also provide results obtained from the Born--Markov quantum master equation (QME) technique for evaluating the FCS.\cite{bagrets_full_2002}
        The QME results show clear steps in the current and the noise when a resonant transport channel opens up (dashed vertical lines), and a flattening of the curves in the high bias regime where the voltage exceeds the bandwidth.
        These features are washed out in data from the NCA-based methods, because the NCA accounts for broadening effects that are neglected in the Markovian QME. The discrepancies between the QME and NCA-based methods arise from two main factors: first, the QME entirely omits Kondo transport, which is expected to occur at the simulated temperature $T=0.5\Gamma \lesssim T_k$ .
        Second, the small bandwidth makes the Markovian approximation inaccurate for these parameters\cite{ridley_lead_2019}.

        The agreement between the NCA and SS-NCA results is excellent at most of the parameters shown.
        However, in the particle--hole asymmetric case and for bias voltages exceeding the bandwidth, we find a discrepancy between the two calculations in the noise (see shaded region of bottom right panel).

		\paragraph*{Slow dynamics.}
		                
		\begin{figure}
			\includegraphics{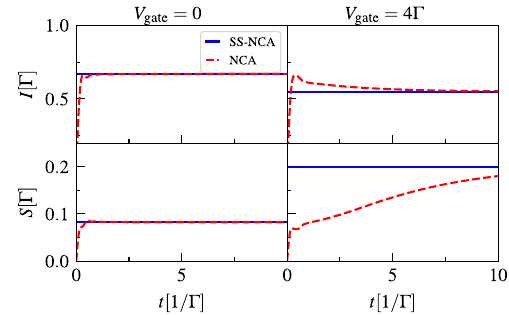}
			\caption{
				\label{fig:dynamics_of_IS}
				Current (top panels) and noise (bottom panels) extracted from the first two cumulants of the FCS obtained from time propagation along with the steady state value from the self-consistent calculation.
				The bias voltage is set to $V=20\Gamma$ in all cases.
				Two gate voltages are shown: $V_{\mathrm{gate}}=0$ (left panels) and $V_{\mathrm{gate}}=4\Gamma$ (right panels).
			}
		\end{figure}
		
\begin{figure}
        	\includegraphics{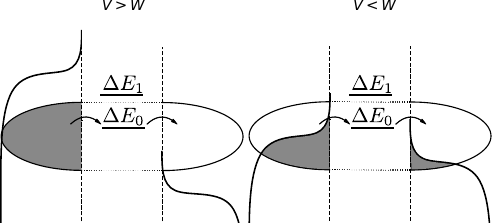}
        	\caption{
        		\label{fig:energy_picture}
        		Simple picture of resonant transport at a bias voltage $V$ that is either larger (left panel) or smaller (right panel) than the bandwidth $W$.
        		The solid lines represent resonance energies $\Delta E_0=\Delta E(V_\text{gate}=0)$ and $\Delta E_1=\Delta E(V_\text{gate}=4 \Gamma)$, occurring at the difference between the energy of two impurity states with occupation differing by one and denoted by the dashed vertical lines in Fig.~\ref{fig:benchmark}.
        		The bands are represented by their coupling densities, with the gray shaded areas indicating populated electronic states.
        	}
        \end{figure}

        In order to understand the apparent disagreement between the NCA and SS-NCA methods, it is useful to examine the time dependence of the current and noise as calculated by the time-dependent NCA method in Fig.~\ref{fig:dynamics_of_IS}.
        We observe that with particle--hole symmetry, the transport current and noise rapidly converge to the value calculated by the steady state scheme at long times.
        However, the convergence is considerably slower for the asymmetric case $V_{\mathrm{gate}}=4\Gamma$, such that the results obtained via propagation are not yet fully converged at time $10/\Gamma$.
        The reason for the slower relaxation in this case has to do with the impurity state energies and the finite bandwidth.
        For $V_{\mathrm{gate}}=4\Gamma$, the system has a resonant transport channel associated with the energy $\Delta E(V_\text{gate}=0)=0$ alongside a second transport channel at $\Delta E(V_\text{gate}=4\Gamma)=8\Gamma$, which is at the band edge (see the dashed vertical lines in Fig.~\ref{fig:benchmark} and the accompanying illustration in Fig.~\ref{fig:energy_picture}).
        Once the bias voltage exceeds the second transport channel, resonant transport through the second resonant channel is energetically allowed, but the weak coupling to the leads at the band edge promotes higher-order effects that slow down the relaxation dynamics.
        This is captured even at the level of the QME, where the transition rate to the second resonance channel is zero while the transition rate to the first channel is one.
        This indicates that there is little ballistic transport through the second resonant channel, and energy transport occurs through a higher-order processes included (approximately) by the NCA scheme.
       
        If we were to simulate time propagation to much longer timescales, we would expect the time-dependent method to reproduce the result given by the steady state scheme, since the level of approximation in the self-energy is identical.
        That would, however, incur a high computational cost.
        This difficulty demonstrates a significant advantage of the SS-NCA method over direct time propagation in parameter regimes exhibiting slow dynamics.
        At lower temperatures, where correlation functions and propagators have a longer range, this problem is generally exacerbated and the SS-NCA becomes increasingly advantageous.
        Nevertheless, treating systems at lower temperatures does generally incur an implicit but clearly increasing cost even within the SS-NCA.
        This is because the time grids must be extended to encompass the full range of dynamical correlation, and because the number of self consistency iterations needed to attain convergence tends to increase in the presence of such long-lived correlations.
        		
        \paragraph*{Scaling function and FCS at steady state.}
        
        The steady state formalism does not provide access to the (diverging) generating function of transport FCS at steady state, but rather to the complex scaling parameter $w(\lambda)$ of Eq.~\eqref{eq:asymptotic_ansatz}, which in turn provides access to the time derivatives of all cumulants (i.e. to the current, noise, etc., see Eq.~\eqref{eq:def_cumulants_steady_state}).
        It is therefore useful to examine our result for $w(\lambda)$ at steady state and compare it to the corresponding time-dependent property.
        
\begin{figure}
	\includegraphics{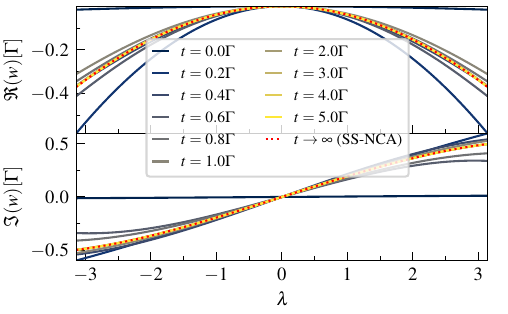}
	\caption{
		\label{fig: generating function}
		The real (top panel) and imaginary (bottom panel) parts of the FCS scaling parameter $w(\lambda)$ at half filling and a bias voltage of $V = 4\Gamma$, with other parameters as in Fig.~\ref{fig:benchmark}.
		The steady state result is plotted as a dashed line, while the finite time results are depicted by solid lines.
	}
\end{figure}

        In Fig.~\ref{fig: generating function} we present the steady state scaling parameter obtained from the SS-NCA equations in the entire range of the counting field, $\lambda\in[-\pi, \pi]$.
        The parameters are as in Fig.~\ref{fig:benchmark}.
        It is easy to show that this should uphold the symmetry $w(\lambda)=w^{\dagger}(-\lambda)$, which is indeed the case to within numerical errors.
        We also evaluate the time dependence of $Z_L\left(t,\lambda\right)$ from NCA, and use Eq.~\eqref{eq:asymptotic_ansatz} to to calculate $w(\lambda)$ at a sequence of finite times.
        At these parameters, the NCA results rapidly converge to the SS-NCA scaling parameter at increasing times.
        This demonstrates that the SS-NCA method is able to efficiently and accurately access the all information contained in the steady state FCS, not just the first few moments.

    \subsection{\label{sec:results} Influence of the lead dimensionality on transport characteristics}
    
    \begin{figure}
    	\includegraphics{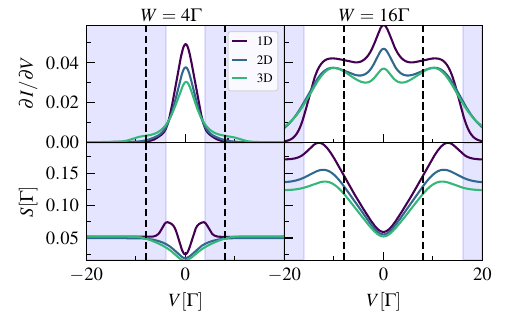}
    	\caption{
    		Influence of lead dimensionality on the conductance (top panels) and noise (bottom panels) at temperature $T=0.5\Gamma$.
    		Results for bandwidths $W=4\Gamma$ (left panels) and $W=16\Gamma$ (right panels) are shown.
    		The dashed vertical lines mark the onset of resonant transport $V=\pm U$, while the shaded region denotes bias voltages that are larger than the bandwidth in the 1D case.
    		\label{fig:dI_S(band)}}
    \end{figure}
    
        We have demonstrated that the SS-NCA method provides results consistent with those obtained from time propagation based on the NCA, though at improved accuracy and reduced cost.
        We therefore now turn to an application.
        In particular, we use the SS-NCA to explore the influence of lead dimensionality, as illustrated in Fig.~\ref{fig:coupling_density_and_systems}, on transport characteristics.
        We also consider two different sets of lead parameters: one where the bandwidth is smaller than the resonant transport threshold ($t_{\text{tb}} = t_{\sigma \mathcal{T}} = \Gamma$, such that in 1D $W=4\Gamma$); and one where the bandwidth spans the resonant and non-resonant transport regimes ($t_{\text{tb}}=4\Gamma$ and $t_{\sigma \mathcal{T}} = 2\Gamma$, such that in 1D $W=16\Gamma$).
        We will focus only on the particle--hole symmetric case ($V_{\text{gate}} = 0\Gamma$) and on a temperature of $T=0.5\Gamma$, except where stated otherwise.
        
        \paragraph*{Kondo physics and lead dimensionality.}
        Fig.~\ref{fig:dI_S(band)} depicts the dependence of the differential conductance (top panels) and noise (bottom panels) on voltage, for the three different lead geometries (1D, 2D and 3D) and two different bandwidths (left and right panels).
        
        For all lead geometries and both bandwidths, we observe  a sharp peak  in the conductance around zero bias voltage.
        At the chosen temperature, we expect the peak to be associated with the Kondo effect.
        At large bandwidth (upper right panel) this low energy enhancement of the transport can be clearly separated from the weakly correlated transport features appearing at higher energies.
                
        In the small bandwidth case $W=4\Gamma$ (left panels in Fig.~\ref{fig:dI_S(band)}), the conductance decays uniformly and in a similar manner with bias voltage in all three lead geometries.
        Higher dimensionality leads to lower, broader features, similarly to what occurs in the coupling density.
        However, the noise in the 1D system (lower left panel) behaves in a manner that is qualitatively different from the 2D and 3D cases.
        Specifically, we note a rapid increase in the noise with voltage, which extends beyond the band edges (i.e. into the shaded region).
        This ``bump'' in the noise characteristics eventually disappears with increasing bias voltage.
        The lack of the feature in the 2D and 3D cases suggests that it is related to the sharp 1D band edge.
        We remark that within the feature the system is characterized by slow relaxation dynamics, for reasons analogous to those discussed in Sec.~\ref{sec:benchmark}; it would therefore be difficult to resolve accurately using a time-dependent method.

		When we instead consider larger bandwidths $W=16\Gamma$ (right panels in Fig.~\ref{fig:dI_S(band)}), the conductance and noise are qualitatively the same for all three lead geometries.
		Features similar to the  previously observed``bump'' appear in the noise characteristics at all dimensionalities, but here they appear at voltages still \emph{within} the bandwidth and concurrent with the appearance of charge sidebands in the conductance.

        \begin{figure}
            \includegraphics{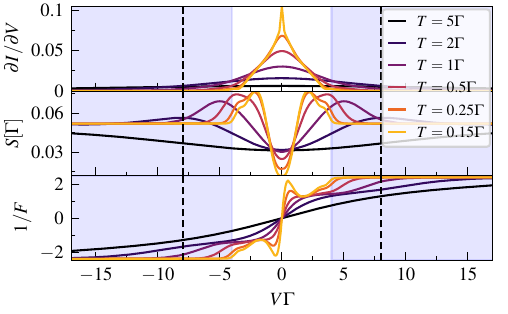}
            \caption{
                Temperature dependence of the conductance (top panel), the noise (middle panel), and the inverse Fano factor (bottom panel) for an impurity coupled to 1D leads with a narrow bandwidth of $W=4\Gamma$. 
                The dashed vertical lines mark the onset of resonant transport $V= \pm U$, while blue shaded areas mark bias voltages that are larger than the bandwidth.
                \label{fig:dI_S(T)}
                }
        \end{figure}
        
		\paragraph*{Temperature dependence.}
		We now consider the effect of temperature on the unusual voltage dependence of the noise in the 1D case.
		Fig.~\ref{fig:dI_S(T)} shows the conductance (top panel) and noise (middle panel).
		The conductance exhibits a Kondo peak that appears at low temperatures, along with a reduction of the temperature-driven broadening at the band edge.
		The noise exhibits the bump feature that was previously observed, which becomes increasingly prominent at the same range of temperatures, suggesting at least some connection to Kondo physics.
		Moreover, at decreasing temperatures, the bump shifts towards lower energies and becomes narrower.
		At the lowest two temperatures shown, the bump is below the band edge, but a second, smaller bump becomes visible closer to the edge.
		
		We associate the first enhancement in noise with the shift from Kondo transport to noisy sequential tunneling through the charge states, and the second with an overlap between the shifted Kondo peaks and the sharp band edges.
		These features overlay a general decrease of the noise towards the high-temperature limit as the voltage increases.
		
		In transport experiments, it is often easier to access the Fano factor $F=S/I$ rather than the noise and current themselves.
		The two-bump structure predicted here can also be seen in the inverse Fano factor, $1/F=I/S$, which we plot in the bottom panel of Fig.~\ref{fig:dI_S(T)}.
		Because the first noise bump occurs in a regime where the current no longer increases rapidly with voltage, it results in a dip in $1/F$.
		The second bump is too small to cause a dip, but results in a clear shoulder with a distinct signature that should be identifiable in experiments.
        
\section{Conclusions\label{sec:conclusions}}
	We introduced a technique for evaluating the FCS of transport observables directly at the steady state, by considering the asymptotic behavior of the generating function rather than its full dependence on time.
	Using this, we showed how a steady state equivalent of the propagator NCA method---which we called the SS-NCA---can be implemented for the nonequilibrium Anderson impurity model, and benchmarked it against the results obtained from the standard NCA using time propagation.
    The SS-NCA capitalizes on reducing the two-time structures of restricted propagators on the Keldysh contour to a time difference representation that becomes exact in the steady state. 
    This reduction enables evaluations at a computational cost that scales approximately linearly with the system's coherence time.
    The numerical scheme includes a self consistency condition for the restricted propagators, which is solved iteratively.
    
    The SS-NCA strikes a useful balance between computational cost, complexity, and accuracy.  
    Like its time-propagation-based counterpart, it can qualitatively capture some strong correlation physics, but at greatly reduced cost, and we showed how important this becomes in the presence of slow relaxation dynamics.
    We then applied the method to the AIM with leads having different dimensionalities, showing that the noise response differs significantly near the Kondo regime.
    This demonstrates that FCS can be utilized as a sensitive probe of a quantum junction's environment.
    
    Looking forward, the SS-NCA can be efficiently applied to a wide variety of quantum impurity problems in and out of equilibrium.
    Moreover, the asymptotic form of the generating function can be used to construct a wide variety of more advanced approximations.
    Future work will include its application to inchworm Monte Carlo methods, where we envision it providing numerically exact access to FCS in the steady states of strongly correlated quantum impurity models.

\section*{Acknowledgements}
    This research was supported by the ISRAEL SCIENCE FOUNDATION (Grants No. 2902/21 and 218/19) and by the PAZY foundation (Grant No. 318/78). 
    A.E. and E.G. were supported by the U.S. Department of Energy, Office of Science, Office of Advanced Scientific Computing Research and Office of Basic Energy Sciences, Scientific Discovery through Advanced Computing (SciDAC) program under Award Number DE-SC0022088. This research used resources of the National Energy Research Scientific Computing Center, a DOE Office of Science User Facility supported by the Office of Science of the U.S. Department of Energy under Contract No. DE-AC02-05CH11231 using NERSC award BES-ERCAP0021805.
    The collaborative work between the US and Israel was enabled by NSF-BSF grant no.~2023720.

\appendix
\section{Computational scaling of the steady state formulation of the full counting statistics \label{app:scaling}}

\begin{figure}
        \includegraphics{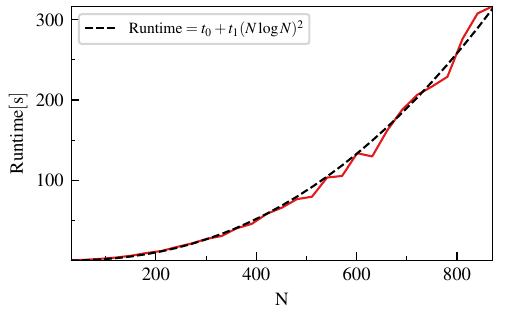}
        \includegraphics{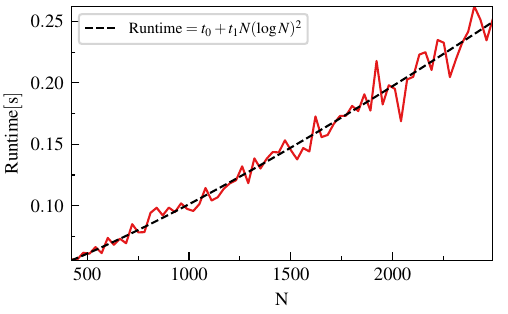}
        \caption{
        	Comparison of the computational cost for accessing the FCS in the steady state by time propagation (top panel) and directly thought the steady state formulation (bottom panel) as a function of the number of points $N$ on the time grid. The black dashed lines represent a fit of the theoretical expectation to the actual computational cost.
        	The parameters used here are $T=5\Gamma$, $V=4\Gamma$, $V_{\text{gate}}=0\Gamma$ and $U=8\Gamma$, with 1D leads and with $t_{\sigma \mathcal{T}}^2 = t_{\text{tb}} = 16\Gamma$.
        	}
        \label{fig:runtime}
    \end{figure}

    As discussed in Sec.~\ref{sec:scaling}, the steady state formulation of the FCS scales more favorably with the number of points $N$ on a (uniform) time grid, when compared to approaches based on time propagation.
    Fig.~\ref{fig:runtime} compares the runtime for accessing the FCS in the steady state by propagation and through the steady state approach for a representative set of parameters.
    As outlined in Sec.~\ref{sec:scaling}, we expected a scaling of $\mathcal{O}(N\log(N))$ for the steady state formulation, 
    and a scaling of $O(N^{2}\log(N))$ when propagating in time.
    The measured runtime of our implementation numerically confirms this theoretical expectation.
    
    The steady state formulation relies on a self consistent iteration scheme, and we found that the typical number of iterations required for convergence is $\sim10$ at the parameters used in Fig.~\ref{fig:runtime}.
    This is accounted for in our measurements, but it remains important to note that the number of iterations needed to achieve convergence generally depends on the parameters and the initial guess.
\bibliography{refs.bib}

\end{document}